# Testing the Validity of the Lorentz Factor

Main Authors: H Broomfield[1], J Hirst[1], Theodoros Vafeiadis[2], Markus Joos[3]
A Singh[1], M Raven[1], T K Chung[1], J Harrow[1], T Kwok[1], J Li[1], K Tsui[1], A Tsui[1], R Perkins[1], H Mandelstam[1], D Khoo[1], J Southwell[1], J Martin-Halls[1], D Townsend[1], H Watson[1].

- [1] Colchester Royal Grammar School, United Kingdom
- [2] Aristotle University of Thessaloniki
- [3] CERN

## Abstract
Our proposed experiment [1] aimed to test the validity of the Lorentz factor with two methods: The time of flight (TOF) of various particles at different momenta and the decay rate of pions at different momenta. Due to the high sensitivity required for the second method the results were inconclusive, therefore we report only on the results of the first method.

## Introduction
This experiment was made possible by the 2016 CERN Beamline for Schools Competition [2]. Our team, named 'Relatively Special', was lucky enough to win this global event, along with 'Pyramid Hunters' from Poland, to design an experiment to carry out at the T9 facility at CERN. We are 17 secondary school students aged 16-18 from Colchester Royal Grammar School, Colchester, United Kingdom who proposed an experiment to test the validity of the Lorentz factor by measuring the time of flights of particles at different momenta [1].

## T9 Beamline
The experimental area encompasses an area of 5 m by 12 m in which different detectors [3], outlined below, can be positioned. The beam entering the experimental area is composed of positive or negative particles. The positive beam contains protons, pions, kaons and positrons and the negative contains their respective antiparticles. Figure 1 shows estimations for the respective proportions of particles at each momentum. The beam momentum could be set between 0.5 and 10 GeV, delivering bursts of roughly $10^6$ particles in 0.4 s.

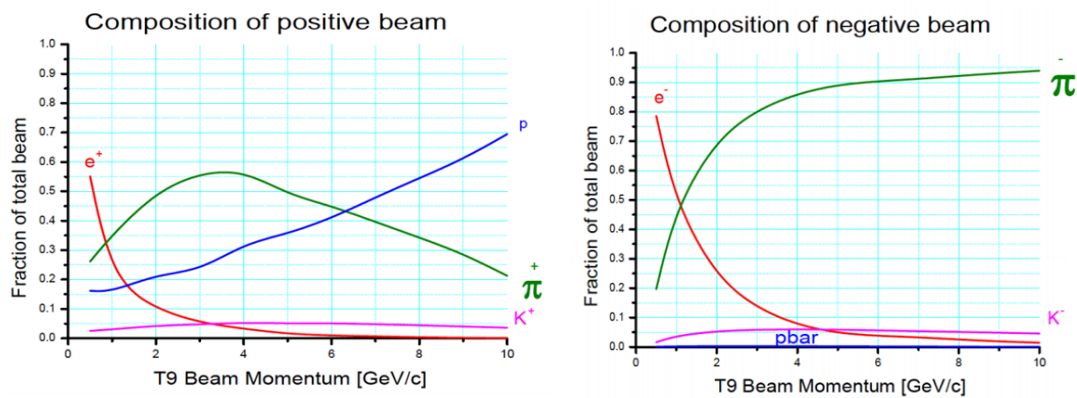

*Figure 1: Composition of the beam in T9.*

# Method

The Lorentz factor, γ, describes the relativistic change in properties of an object that is moving with respect to an inertial frame at velocity, v. As particles travel close to the speed of light, c, in particle accelerators, the effects of special relativity become significant enough to measure. This includes the effect of mass increase.

$$\gamma = \frac{1}{\sqrt{1-\frac{v^2}{c^2}}} \quad \text{Equation 1}$$

$$p = \gamma m_0 v \quad \text{Equation 2}$$

Equation 2 describes the relativistic momentum, p, of a particle with rest mass, $m_0$.

From the previous two equations using the time, t, to travel a distance, d, in the inertial frame's perspective, equation 3 can be derived:

$$t = \frac{d\sqrt{m_0^2 c^2 + p^2}}{pc} \quad \text{Equation 3}$$

Using the facility at CERN, a beam of particles at a certain momentum could be defined, and the time it took for them to travel a known distance could be measured. These results could then be compared with the theoretical predictions given by equation 3 and seen in Figure 2 for all particles

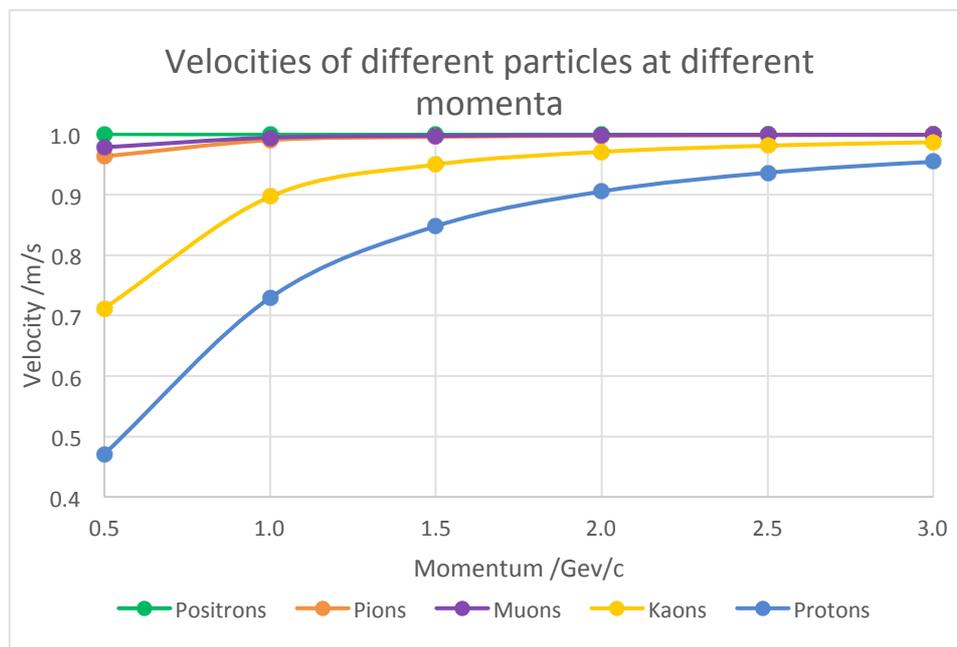

Figure 2: Velocity of the particles composing the T9 beam.

of the beam.

## Apparatus

The experimental set up is outlined in Figure 3 and Figure 4. This allowed us to record the time of flight (TOF) of different particles within the beam.

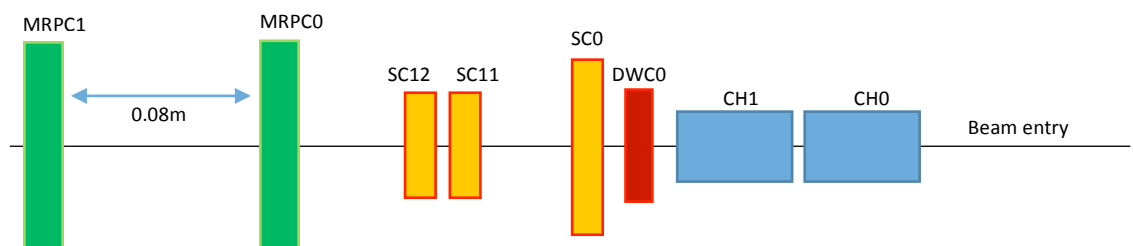

Figure 3: TOF first setup.

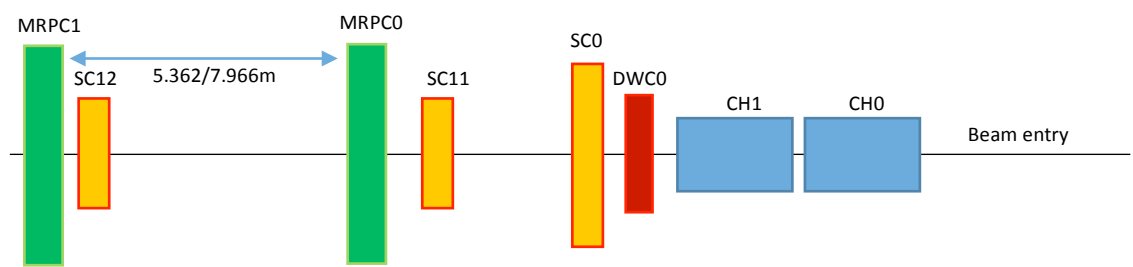

Figure 4: TOF second setup.

### Cherenkov Counters (CH0, 1)

Two Cherenkov detectors are part of the fixed detectors in T9. They can be tuned to discriminate between electrons, muons and pions by appropriately changing the pressure of the gas inside the detectors. The identification of heavier particles is not possible in T9 for technical reasons.

The Cherenkov's, filled with $CO_2$, were used to tag only the electrons so we could find an exact time of flight for each of the distances. Using these times, we could work out the effective lengths allowing errors in measurements to be accounted for.

### Scintillators (SC0, 11, 12)

Beamline for Schools has a large number of scintillators of various shapes and sizes. In addition, one scintillator is part of the fixed detectors in T9.

Three scintillators were used to reduce the chance of a false signal. This was facilitated by NIM modules which were used to connect the detectors to ensure that a particle would only be registered if it passed through all three (and the two MRPCs) in an event known as a coincidence.

Any background noise would be filtered out by the unlikely chance that a random signal would appear in all at the right times.

SC11 and SC12 were 1x1cm$^2$ which allowed a very narrow beam to be defined. Although using a narrow beam significantly reduced the rate, it facilitated the analysis of the data.

**Multi-gap Resistive Plate Chamber (MRPC0, 1)**

The Multi-gap Resistive Plate Chambers (MRPC) [4] are gaseous detectors. The MRPCs of Beamline for Schools consist of six equally spaced resistive plates and have an active area of 30 x 30 cm$^2$. The signal is connected from both sides of the 24 strips.

They can provide very accurate time information for the passage of a particle therefore they were used to measure the time of flight of the particles. Two MRPCs were positioned at a distance to allow us to calculate theoretical predictions for the time of flights.

**Experimental Procedure**

*5.362 m*

Our first set up, Figure 3, was with the MPRCs at a separation of 5.362 m. This distance was selected as it was far enough for the results to have a noticeable change in times as we increased momentum from 0.5 GeV/c to 2.5 GeV/c in 0.5 GeV/c intervals. Our lowest momentum was limited by the beam composition and the highest by the fact that at large momenta, the speeds of particles tend to that of light and become very close until our equipment could not distinguish them.

*7.966 m*

The next set up had the MRPCs 7.966 m apart, which was the maximum possible distance we could achieve in the experimental area. We also moved SC12 so it was positioned just in front of MPRC1 so that the beam width was kept constant during the runs, seen in Figure 4. Although this lowered the count rate, it increased accuracy of the data. Runs were taken from the range of 1 GeV/c to 3 GeV/c in 0.5 GeV/c intervals. The reading at 0.5 GeV/c was excluded as the proportion of protons was too low and included at 3 GeV/c because the greater distance meant distinguishing particles at higher velocities was possible.

# Analysis

The time of arrival of the particles to each MRPC was calculated as the average time that was recorded by the cards on both sides of the strip that was hit. The time of flight was calculated by subtracting the time of arrival of the first MRPC from the time of arrival of the second.

TOF measurements, precise in a picosecond level, would require a careful alignment and positioning of the detectors, study of the timings of the electronic chain and the systematics of the experimental setup. Although we isolated the major parameters such as cable lengths and timing of the signals, it became evident that deeper study that exceeds the scope of the competition was required.

Therefore, a method was developed to allow us to factor in the systematics without the need to isolate and quantify them. Using runs with 1 GeV/c electrons only we measured their TOF for each of the two setups and calculated an effective length ($L_{eff}$) which would then be used to calculate the theoretically expected TOF for the rest of the particles. The effective lengths for the two distances are seen in Table 1.

# Results

The histograms obtained for each run mostly contained two distinct peaks, one for electrons, muons and pions and another for protons. The masses of the particles in the first peak are similar enough for their TOFs to all be very close over the momentum range (Figure 5). For this reason, and the limited resolution of the equipment, this group of particles was treated as a single, mixed peak.

*Figure 5: A characteristic TOF spectrum (2.5 GeV/c). The left peak corresponds to electrons, muons and pions and the right to protons. Kaons are too few to be identified.*

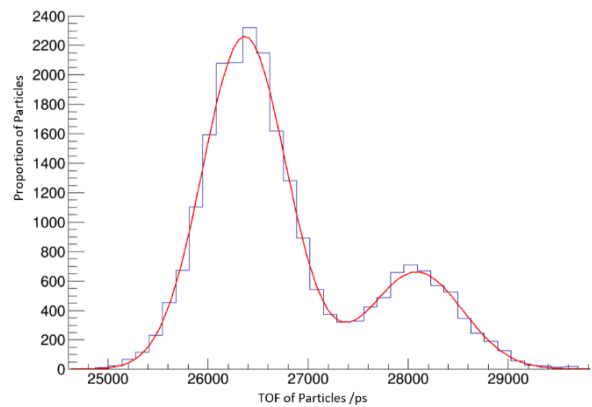

Dual Gaussian distributions were plotted as in Figure 5. This enabled us to find the mean TOF and standard deviation at each momentum for the protons and mixed peaks. Error bars were placed at ±1SD on each of our measured times as seen in Figure 6 and 7.

*The theoretically expected time of flight for all the particles in the two distances are seen in*

Table 1.

| Distance | 7.966 | | | | |
|---|---|---|---|---|---|
| Momentum | TOF_positrons | TOF_muons | TOF_pions | TOF_kaons | TOF_protons |
| 1.0 | 26553 | 26699 | 26812 | 29605 | 36407 |
| 1.5 | 26553 | 26618 | 26669 | 27951 | 31318 |
| 2.0 | 26553 | 26590 | 26618 | 27348 | 29329 |
| 2.5 | 26553 | 26577 | 26595 | 27065 | 28361 |
| 3.0 | 26553 | 26570 | 26582 | 26909 | 27821 |
| Distance | 5.362 | | | | |
| Momentum | TOF_positrons | TOF_muons | TOF_pions | TOF_kaons | TOF_protons |
| 0.5 | 17873 | 18263 | 18561 | 25100 | 37997 |
| 1.0 | 17873 | 17972 | 18048 | 19927 | 24506 |
| 1.5 | 17873 | 17917 | 17951 | 18814 | 21080 |
| 2.0 | 17873 | 17898 | 17917 | 18408 | 19741 |
| 2.5 | 17873 | 17889 | 17901 | 18218 | 19090 |

*Table 1: Theoretically expected TOF for all the particles of the beam for the two distances.*

Our results are seen in **Error! Reference source not found.** (5.362 m) and (7.966 m).

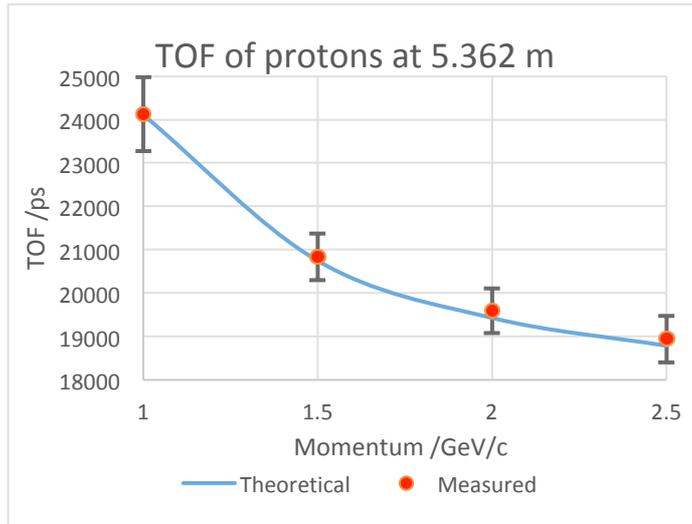

*Figure 7 : TOF of protons at 5.362 m. The blue line corresponds to the theoretically predicted values and the red dots to the measured ones.*

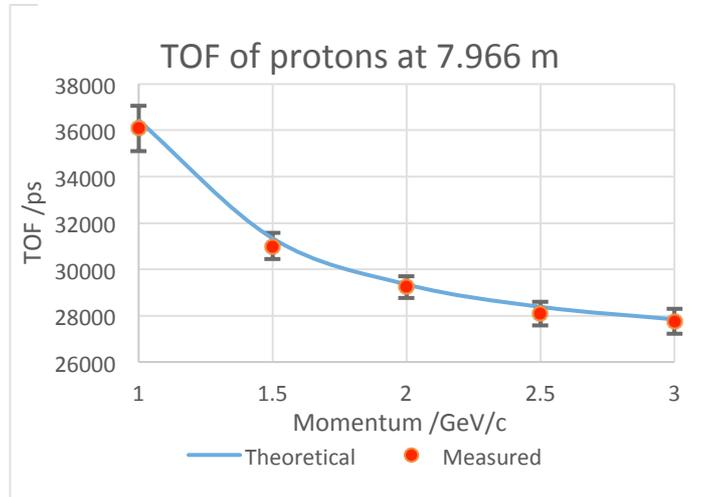

*Figure 6: TOF of protons at 7,966 m. The blue line corresponds to the theoretically predicted values and the red dots to the measured ones.*

# Conclusion

The trend of the experimental values appears to match closely with the theoretical predications as shown in Figure 7 and Figure 6. We can therefore say that our results verify the relativistic effect of the Lorentz factor.

# Acknowledgements

We would like to thank CERN for providing us with a wholly educational and enjoyable experience [5] by making our experiment possible. We would also like to thank the organisers of the competition and the scientists who volunteered their time for us and without whom would have made this impossible: Mr Markus Joos, Dr Theodoros Vafeiadis, Dr Alexander Hristov, Dr Oskar Wyszynski, and many more.

The underlying experiment was supported by the CERN & Society Foundation, funded in part by the Arconic Foundation, with additional contributions from the Motorola Solutions Foundation, the Ernest Solvay Fund managed by the King Baudouin Foundation, as well as National Instruments.

# References

[1] Proposal from Relatively Special and Video

[2] http://beamline-for-schools.web.cern.ch/

[3] Beamline for Schools – Beam and Detectors

[4] ALICE -Technical Proposal for A Large Ion Collider Experiment at the CERN LHC, CERN/LHCC/95-71, December 1995.

[5] Relatively Special CERN video

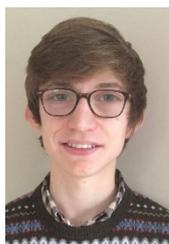

**H Broomfield** was a student at Colchester Royal Grammar School and is studying Mathematics with Physics at Cambridge University.

**J Hirst** was a student at Colchester Royal Grammar School and is going to study Mathematics at Cambridge University.

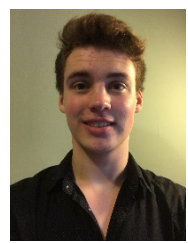

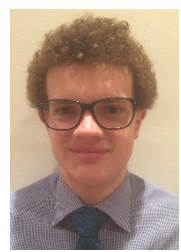

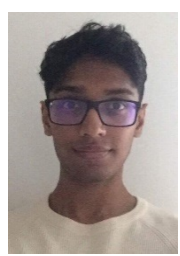

**A Singh** was a student at Colchester Royal Grammar School and is studying Mathematics at Imperial College London.

**M Raven** was a student at Colchester Royal Grammar School and is studying at Massachusetts Institute of Technology.

**T K Chung** was a student at Colchester Royal Grammar School and is currently an Engineering undergraduate at the University of Cambridge.

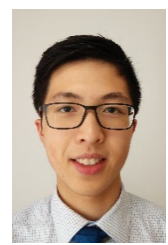

**J Harrow** was formerly a student at Colchester Royal Grammar School and is now going to study Mathematics and Physics within Natural Sciences at Durham University.

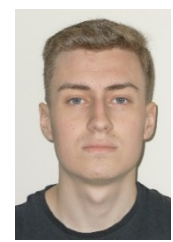

**T Kwok** was a student at Colchester Royal Grammar School and is studying Natural

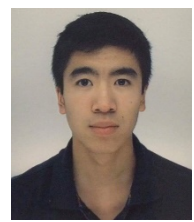

Sciences at Magdalene College Cambridge.

**J Li** was a student at Colchester Royal Grammar School and is currently studying Medicine at the University of Cambridge.

**K Tsui** was a student at Colchester Royal Grammar School and is studying Dentistry at the University of Hong Kong.

**A Tsui** was a student and a boarder at Colchester Royal Grammar School and is studying Medicine at Imperial College London.

**R Perkins**, a past student of Colchester Royal Grammar School, is going to study Physics at Imperial College London.

**H Mandelstam** is a student at Colchester Royal Grammar School and has applied to study Physics at University.

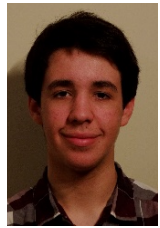

**J Southwell** was a student at Colchester Royal Grammar School and is studying Mechanical Engineering at Bath University.

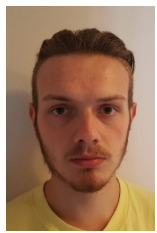

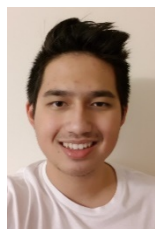

**D Khoo** was a student at Colchester Royal Grammar School and is currently studying at the University of Bath and is working towards a degree in Mechanical Engineering.

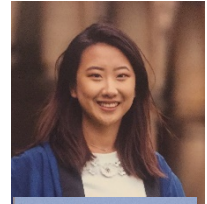

**J Martin-Halls** was a student at Colchester Royal Grammar School and is currently working as a Consulting Engineer, intending to study Engineering at university.

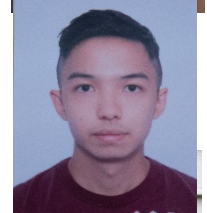

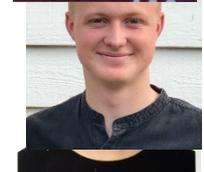

**D Townsend** is a student at Colchester Royal Grammar School and has applied to study Mathematics at University.

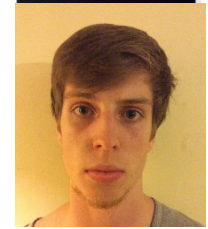

**H Watson** is a student at Colchester Royal Grammar School. He intends to study Mathematics at university.

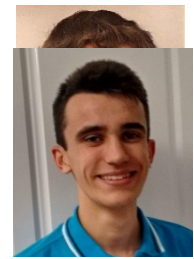